\documentclass[11pt,letterpaper,onecolumn]{article}
\usepackage{indentfirst}
\usepackage{url}
\usepackage[top=2cm,bottom=2cm]{geometry}
\usepackage{comment}
 
\newcommand{\p}{\partial}

\newcommand{\Lag}{\mathcal{L}}

\usepackage{answers}
\usepackage{authblk}
\usepackage{cancel}
\usepackage{setspace}
\usepackage{blindtext, rotating}
\usepackage{graphicx}
\usepackage{enumitem}
\usepackage{textcomp}
\usepackage{multicol}
\usepackage{mathrsfs}
\usepackage{amsmath,amsthm,amssymb}
\usepackage{indentfirst}
\usepackage{gensymb}
\usepackage{mdframed}
\usepackage{tikz-cd}
\usepackage{fancyhdr}
\usepackage{tensor}
\usepackage{braket}
\usepackage{float}
\title{\bf Generalized multi-dimensional conservation laws for stimulated Raman and 
Brillouin scattering in a density gradient
}

\author[1]{Vijay Patel}
\author[1]{Sarah Chase}
\author[1]{Frank S. Tsung}
\author[3]{John P. Palastro}
\author[4]{Denise E. Hinkel}
\author[1,2]{Warren B. Mori}
\affil[1]{Department of Physics and Astronomy, University of California, Los Angeles, California 90095, USA}
\affil[2]{Department of Electrical and Computer Engineering, University of California, Los Angeles, California 90095, USA}
\affil[3]{University of Rochester, Laboratory for Laser Energetics, Rochester, New York 14623, USA}
\affil[4]{Lawrence Livermore National Laboratory, Livermore, California 94550 , USA}

\begin{document}
\maketitle

\begin{abstract}
Generalized local and multi-dimensional conservation laws 
of 
action, energy, momentum, and angular momentum are derived for 
stimulated Raman (SRS) and Brillouin backscattering (SBS) in a density gradient
within the paraxial ray approximation. A Lagrangian density is given for which requiring the action be stationary
reproduces the well known envelope equations for SRS and SBS in density 
gradients in the absence of damping. Using Noether's theorem, the  symmetries of the 
Lagrangian density are used to obtain local conservation laws for quantities 
that can easily be identified as the wave action, energy,  and momentum. These multi-dimensional  conservation laws reduce 
to the well known one dimensional Manley-Rowe relations, and frequency and wavenumber 
matching conditions. Additional symmetries of the action lead to conservation laws for new quantities that are identified as the orbital angular momentum and additional contributions to the energy and momentum of the wave from frequency and wavenumber shifts.
Augmentation of the conservation laws in the presence of damping and extensions of the Lagrangian to include higher order corrections to the paraxial ray equation and nonlinear frequency shifts are also provided. 
\end{abstract}



\section{Introduction}
The nonlinear optics of plasmas (NLOP) is the study of how lasers (electromagnetic waves) nonlinearly interact within a plasma 
NLOP. It is thus a fundamental topic within plasma physics that includes processes such as harmonic generation \cite{esarey2002nonlinear}, four-wave mixing \cite{federici2002review},\cite{lehmberg1978theory}, self-focusing \cite{max1975strong}, filamentation \cite{kaw1973filamentation}\cite{berger1993theory}, and laser plasma instabilities (LPI)
\cite{forslund1975theory, drake1974parametric, kruer1988physics, michel2023introduction}. LPI have been  extensively researched for over half a century primarily motivated by its role in Inertial Fusion Energy (IFE) 
. 
In laser driven IFE \cite{betti2016inertial} there are two distinct concepts, direct \cite{craxton2015direct} and indirect drive \cite{lindl1995development}.
In direct drive, several lasers are fired isotropically upon a target that consists of a shell of material such as plastic and a cryogenically cooled fuel consisting of Deuterium and Tritium. The lasers ablate the outer shell of the fusion pellet, creating a series of delicately timed shock fronts that compress the fuel to ignition regimes \cite{betti2016inertial}\cite{craxton2015direct}. On the other hand in indirect drive, the target is placed in a container known as a hohlraum with a high Z material on the inner walls. 
Lasers are then directed towards the walls through small openings in the hohlraum. The walls of the hohlraum absorb the laser light and re-emit the energy in the form of X-rays, providing an isotropic bath of radiation to compress the target. Besides needing to be isotropic, the pressure (temperature) from (
the x-rays must follow a carefully designed temporal profile to launch a series of shocks to compress the fuel\cite{betti2016inertial}\cite{lindl1995development}.

In both 
direct and indirect drive, the lasers generally propagate through a low density plasma (below the critical density, $n_c\equiv \omega_0^2 \frac{m}{4\pi e^2}$ where $\omega_0$ is the laser frequency). In the case of direct drive, the plasma 
is the corona generated from the ablation and blowoff of the shell, while in indirect drive it is from the ionization of a low density gas fill inserted into the hohlraum to prevent the blowoff from the walls from crossing the path of the incoming lasers. 

As the lasers propagate through the relatively long plasmas they are susceptible to a range of LPI processes \cite{kruer1988physics}\cite{michel2023introduction} including Stimulated Raman Scattering (SRS)\cite{forslund1975theory}\cite{drake1974parametric}, Stimulated Brillouin Scattering (SBS)\cite{forslund1975theory}\cite{drake1974parametric}, cross beam energy transfer (CBET) \cite{KruerCBET}\cite{ mckinstrie1996two}\cite{michel2009energy}, and two plasmon decay (2$\omega_p$ or TPD)\cite{liu1976parametric}\cite{afeyan1995unified}\cite{simon1983inhomogeneous}\cite{LangdonTPD}.  Several of these processes, e.g., SRS and SBS backscatter, (BSRS and BSBS)and TPD can be described as three-wave parametric instabilities  in which the incident laser light (pump) decays into two other waves (daughter waves) mediated by the ponderomotive force in the plasma \cite{kruer1988physics}\cite{michel2023introduction}. In BSRS the laser decays into an electron plasma wave (EPW) and a backscattered electromagnetic wave (EM), in BSBS it decays  into an ion acoustic wave (IAW) and a backscattered EM wave, and in TPD it decays into two sideways propagating EPWs. These processes obey frequency and wave number matching conditions that are related to energy and momentum conservation.  In SRS and SBS the daugther (EPW/IAW and backscattered EM) waves can propagate at angles with respect to the incident light and this is referred to as sidescatter \cite{Liu1973}. Forward scatter \cite{forslund1975theory}\cite{drake1974parametric} is also possible; however, in this case there can be four waves, the pump, and EPW/IAW and two EM waves 
. 
%
The growth rates of these LPI depend on the scattering geometry. For instance, scattering processes involving  daughter waves traveling in the opposite direction of the pump (backscatter) produce a much higher growth rate than forward scatter. When daughter waves propagate in opposite directions, there can also be absolute instabilities \cite{forslund1975theory}\cite{Briggs}\cite{Bersabsolute}. 
Furthermore, in BSRS and TPD the  EPWs can lead  to the generation of non thermal electrons\cite{estabrook1980heating}\cite{forslund1975plasma}\cite{LangdonTPD}. 
In the presence of a density gradient, side-scatter geometries tend to be favored in which the daughter light-wave travels predominantly parallel to the density isosurfaces\cite{Liu1973}.


{\indent} In either direct and indirect drive, the occurrence of SRS and SBS can result 
in a reduction and non-uniformity of  the drive,
thereby greatly reducing the compression of the fuel.  Moreover, the resulting plasma modes excited by SRS and TPD 
can produce a supra-thermal band of electrons, which can attain temperatures more than an order of magnitude higher than the initial plasma temperature \cite{estabrook1980heating}\cite{forslund1975plasma}\cite{LangdonTPD} \cite{michel2023introduction}. These hot electrons 
can pre-heat the target which reduces its compressibility. 
Understanding and mitigating
three-wave instabilities
has therefore comprised a major part of previous and current IFE research.  Numerous techniques to control these instabilities include the use of multi-speckled beams (initially developed to mitigate hydrodynamic instabilities by smoothing the drive profile on the target), including by adding bandwidth,  through techniques such as Smoothing through Spectral Dispersion (SSD) \cite{skupsky1989improved}, Induced Spatial Incoherence (ISI)\cite{lehmberg1987theory}, and Spike Trains with Uneven Duration and Delay (STUD)\cite{afeyan2013optimal},  and  
by applying a uniform B-field \cite{bjwin} have been studied experimentally and/or numerically. 
Multi-speckle beams are inherently multi-dimensional and thus theoretical and simulation models for such beams should include three-dimensional (3D) effects. Furthermore, the plasmas of interest are inhomogeneous so that
frequency and wavenumber matching can only be satisfied at one location; thus theoretical and simulation models must include the effect of inhomogeneity\cite{rosenbluth1972parametric}.



\indent The theories for SRS, SBS, and TPD in homogeneous plasmas are well known, and methods for extending the analysis for plane waves in non-uniform plasmas are mature \cite{rosenbluth1972parametric}\cite{liu1976parametric}\cite{afeyan1997vol1variational}.
In density gradients for lasers moving normal to the density gradient,  methods for determining the convective gain and absolute instability threshold are also well known \cite{rosenbluth1972parametric}\cite{liu1976parametric}\cite{simon1983inhomogeneous}\cite{afeyan1997vol1variational}. These methods generally reduce the problem to 1D ODEs in Fourier space for the wave number along the density gradient.  
At sufficiently low density it is often the case that envelope or eikonal approximations are used in both uniform and nonuniform plasmas.   At densities 
much less than the quarter critical density, $n\ll .25n_0/n_c$, SRS and SBS are the  instabilities of importance. And in simple inhomogeneous plasma density profiles these instabilities are convective for densities well below quarter-critical\cite{rosenbluth1972parametric}. 

While the theory for SRS and SBS for plane waves is known,  extending it to multi-speckle beams and beams with bandwidth remains challenging. As a result, numerical simulation of the relevant equations is typically utilized. Some 
simulation tools utilize the envelope approximation in both space and time to obtain three-wave coupled equations. In addition, these equations are 
often based on the paraxial ray approximation. For example, the code p3FD is based on the envelope and paraxial ray approximation for the pump and two daughter waves, and the equations permit simulating density gradients. In one dimension it is well known that the three wave equations obey conservation laws for action, energy and axial (along the laser propagation direction) momentum. Conservation of wave action for three wave interactions is referred to as the Manley-Rowe relations \cite{manley2007some}\cite{forslund1975theory}\cite{kruer1988physics}\cite{michel2023introduction}. 

On the other hand, 
the generalization  of the conservation of wave action (and energy and momentum) to multi-dimensional wave couplings in the paraxial-ray approximation remains to be seen in the literature. In this paper, we present a Lagrangian density whose Euler-Lagrange equations generate the 3D  3-wave coupled equations in density gradients. Lagrangian and variational principles have been shown to be useful for obtaining LPI thresholds near quarter-critical\cite{afeyan1997vol2variational} and for identifying instabilities\cite{duda2000variational} and conservation laws in short-pulse LPI\cite{decker1996evolution}. 
These conservation laws, based on Noether's theorem \cite{noether1918invariante}\cite{noethers} for symmetries of the Lagrangian, can be identified as conservation of action, energy, momentum, and orbital angular momentum (OAM)\cite{allen1992orbital}. OAM of a wave is based on the angular momentum associated with a waves momentum flux. This is described in more detail below.

The connection of these conservation laws to frequency, wavenumber and OAM
matching is discussed.  We obtain two different conservation laws for energy and momentum. One form for energy (and momentum) conservation is not strictly independent from the action,  while the other is in a form that to our knowledge has not been mentioned in the literature before. These conservation laws are useful for understanding how instabilities saturate in 3D and the limits to the amount of light scattered in 3D. They can also be used to check the numerics of algorithms in codes such as pF3D\cite{berger1998dominant}. We note that the starting envelope equations often include phenomenological damping terms. Including dissipation is not amenable to a Lagrangian description. However, we show that if the Euler-Lagrange equations are augmented with damping terms,  then it is straightforward to modify the conservation laws 
to include loss or 
sink terms. We note that Noether's theorem was previously applied to a 1D Lagrangian for the full wave quantities (no envelopes)
to obtain the Manley-Rowe relations of three-wave instabilities in the absence of the Eikonal approximation \cite{brizard1995local}. It was also applied to a 1D Lagrangian for short-pulse LPI\cite{decker1996evolution}.

\indent Although not pursued in this paper, obtaining a Lagrangian might offer some opportunities for new theoretical directions.  In the presence of a density gradient, the coupled-mode equations are typically solved using WKB methods \cite{rosenbluth1972parametric}. Assuming a constant pump 
allows the two equations to be reduced to a second order parabolic ODE which can be solved by patching solutions solved outside of the classical turning points. However, such methods must be fine-tuned depending on the specific profiles used for the density gradient. A variational approach has the advantage of being able to solve for the bound state solutions and absolute growth rates independent of the location of the turning points relative to the interaction region. The variational principle involves minimizing some bi-linear functional of a trial wavefunction by adjusting its corresponding free parameters. The growth rate corresponds to the effective action evaluated for the resulting ground-state wavefunction 
. 
Such a method was employed by \cite{afeyan1997vol2variational} for Gaussian trial wavefunctions. 

Another possible use for a 3D Lagrangian is to substitute trial functions for finite width pumps and daughter waves into the action $W_{12}=\int_{2}^1 dtd\mathbf {x} L$ and carry out the integration in the directions transverse to the propagation direction of the waves.  The trial functions could be parameterized by spot size, radius of curvature, phase and amplitude of envelope solutions. Variation of the reduced 1D Lagrangian leads to differential equations for the parameters of the trial functions. This has been applied to a Lagrangian that describes short-pulse LPI \cite{duda2000variational}. 

\subsection{Starting Equations}

The starting point is the well known and accepted multi-dimensional three-wave envelope equations that describe SRS and SBS backscatter in a density gradient within the paraxial wave approximation \cite{berger1998dominant}, 
\begin{align}
  \omega_0(\partial_t + \nu_0+v_{g0} \partial_z  -\frac{iu_0^2}{2\omega_0}\nabla_{\perp}^2 - 
  \frac{\partial_z v_{g0}}{2} )\tilde{b}_0 &= -ic_0 \tilde{b}_1
  \tilde{b}_2e^{-i\phi(z)} \label{eq: 0}\\
    \omega_1(\partial_t + \nu_1 +v_{g1}\partial_z)  -\frac{iu_1^2}{2\omega_1}\nabla_{\perp}^2-
    \frac{\partial_z v_{g1}}{2} )\tilde{b}_1 &= -ic_1 \tilde{b}_0
    \tilde{b_2}^* e^{i\phi(z)} \label{eq:1} \\
    \omega_2(\partial_t + \nu_2 +v_{g2} \partial_z  -\frac{iu_2^2}{2\omega_2}\nabla_{\perp}^2 - 
    \frac{\partial_z v_{g2}}{2} )\tilde{b}_2 &= -ic_2 \tilde{b}_0\tilde{b}_1^*e^{i\phi(z)} \label{eq:2}\\
  \text{and}  \ 
  \phi(z) \equiv \int dz&(
  k_1(z) + 
  k_2(z) - 
  k_0(z))\nonumber
\end{align}
where the subscripts 0,1, and 2 correspond to the incident light wave- (pump), backscattered light wave,  and forward propagating electrostatic wave (and electron plasma wave-EPW- for SRS and ion acoustic wave-IAW- for SBS) respectively. 
The full complex wave amplitude can be obtained by multiplying each envelope with the corresponding phase so that $b_j = \tilde{b}_je^{i(k_jz - \omega_j t)}$. The coupling constants are $c_0 = c_1 = 1/2$ and $c_2 = c_0  \Gamma_2$ with $\Gamma_2 = c^2k_2^2 / \omega_p^2$. The 1D version of these equations were first described by Rosenbluth (wihtout the $\partial_z v_{g\alpha}$ terms) where the mismatch phase $\phi(z)$ was introduced. The quantities $\omega_j$, $k_j(z) 
$, $\nu_j$, $v_{gj}$, $u_j$, $c_j$, $\tilde{b}_j$, are the frequency, wavenumber, phenomenological damping, group velocity,  wave speed (speed of light, c,  for light waves, $(3k_BT_e/m)^{1/2}
$ for the EPW, and  the sound speed, $c_s=(ZT_e/M)^{1/2}$ for the IAW-where $k_B$ is Boltzman's constant, $T_e$ is the electron temperature, $Z$ is the ionization state of the ion, and m and M are the electron and ion mass respectively),  wave coupling coefficients, and wave envelopes respectively. The equations also assume frequency and wavenumber matching, given in one dimension by $\omega_0 - \omega_1 - \omega_2=0$ and $k_{0,0} - k_{1,0} - k_{2,0}$. The 0 subscript means that the wavenumbers should be evaluated at the resonance point located at $z = 0$. These equations can be combined to model SRS and SBS occurring simultaneously in the absence of hydro motion, and  are the key equations within the code pF3D. 

Although these multi-dimensional equations have been well known for decades, the conservation laws that arise from them have heretofore not been derived.  We show next that a set of conservation laws can be obtained from a Lagrangian density from which the conservation laws can then be obtained. Furthermore, this procedure yields additional conservation laws for energy and momentum that go beyond simple frequency and wavenumber matching.

\section{Lagrangian Formulation}
A set of differential equations that describe the dynamics of a system can often be cast 
in terms of a variational principle, where the action $W_{12}=\int_2^1 dx_\mu\mathcal{L}(b_\alpha,b_\alpha^*,  \partial_\mu b_\alpha, \partial_\mu b_\alpha^*)$ is stationary with respect to variations of the $b_\alpha$ which are functions of the four vector $\mathcal{\zeta
}=(t, z, x)\equiv x^\mu$ with $\mu=0,1,2,3$. The dynamical or Euler-Lagrange (E-L) equations follow from the requirement that $W_{12}$ be stationary to variations about the correct path if the end points are fixed.  It is also well known that incorporating damping into a variational principle can be difficult. Thus, we start by assuming that all the phenomenological damping terms are zero. 

We state without derivation that in the absence of damping that Eqns. 1-3
can be obtained from a Lagrangian density of the form:
\begin{align}
    \mathcal{L} &= 
    \frac{\omega_\alpha}{ic_\alpha}\bigg(\frac{1}{2}(b_\alpha^*\partial_t b_\alpha - b_\alpha \partial_t b_\alpha^*) + (\frac{v_{g\alpha}}{2}(b_\alpha^*\partial_z b_\alpha - b_\alpha \partial_z b_\alpha^*) + 
    \frac{iu_\alpha^2}{2\omega_\alpha}\nabla_\perp b_\alpha^* \cdot \nabla_\perp b_\alpha\bigg)\nonumber\\
    &+ (b_0^*b_1b_2e^{-i\phi(z)} + b_0b_1^*b_2^*e^{i\phi(z)})\label{eq:4}
\end{align}

 where Einstein notation is used so that it is assumed that there is a sum over repeated indices.
It can be easily
demonstrated that the Euler-Lagrange (E-L) equations of the above Lagrangian, 
\begin{align*}
    \partial_t\frac{\p \Lag}{\p\partial_t b_\alpha^*}+\partial_z\frac{\p \Lag}{\p\partial_z b_\alpha^*} +\nabla_\perp \cdot \frac{\p \Lag}{\p\nabla_\perp b_\alpha^*} &= \frac{\p \Lag}{\p b_\alpha^*}, 
\end{align*}

precisely recover Eqns. \ref{eq: 0}-\ref{eq:2}. It is obvious that variations to $b_\alpha$ give the complex conjugate to this equation. We note that the E-L equations can be written more compactly using four vector notation,

\begin{align*}
    \partial_\mu\frac{\p \Lag}{\p\partial_\mu b_\alpha^*}
    = \frac{\p \Lag}{\p b_\alpha^*}
\end{align*}

As just mentioned,  
it is not known 
how to include 
damping terms for each wave into a pure Lagrangian formalism. 
However, for the case of point particles it is widely known that damping can be incorporated by augmenting the Euler Lagrange equations as:
\begin{align*}
    \frac{d}{dt}(\frac{\p L}{\p \dot{q}}) - \frac{\p L}{\p q} &= \frac{\p F}{\p \dot{q}}
\end{align*}
where $F(\dot{q}) = \frac{1}{2}\nu\dot{q}^2$ where $\nu$ is the damping coefficient. In our continuum case,  we can incorporate damping by augmenting the E-L equations with 
$F(b_\alpha, b_\alpha^*, \p_tb_\alpha, \p_t b_\alpha^* ) = \sum_\alpha \nu_{\alpha}\frac{\omega_\alpha}{ic_\alpha}(b_\alpha^*\p_t b_\alpha - b_\alpha \p_t b_\alpha^*)$. We will describe below how this approach will provide a mechanism for augmenting the conservation laws to include loss terms.


\subsection{Conservation laws and global symmetries}
One utility of determining a Lagrangian is that through the identification of internal symmetries local conservation laws can be ascertained. Thus, the challenge is to determine transformations that leave the action invariant. Once these are found, then through Noether's theorem, \cite{noether1918invariante}\cite{noethers} local conservation laws follow. 
\subsubsection{Internal Symmetries}
An internal  symmetry arises  when the action is invariant because the Lagrangian is invariant, $d_\epsilon \mathcal{L}=0$, for transformations of $b_\alpha = b_\alpha(\mathbf{x},t,\epsilon)$ (and $b_\alpha^*(\mathbf{x},t,\epsilon)$). Here, $\epsilon$ parameterizes the symmetry. The variation of $\mathcal{L}$ associated with such a transformation is

\begin{align}
    d_\epsilon \mathcal{L} &= \frac{\p \mathcal{L}}{\p b_\alpha} d_\epsilon b_\alpha + \frac{\p \mathcal{L}}{\p \p_\mu b_\alpha} \p_\mu d_\epsilon b_\alpha + \text{c.c}
\end{align}

where in this case, $d_\epsilon b_\alpha\equiv (db_\alpha/d\epsilon)
\epsilon$.   Setting this to 0 and then integrating by parts leads to, 

\begin{align}
    \p_\mu\bigg(\frac{\p \mathcal{L}}{\p \p_\mu b_\alpha} d_\epsilon b_\alpha\bigg) + \bigg(\frac{\p \mathcal{L}}{\p b_\alpha} - \p_\mu \frac{\p \mathcal{L}}{\p \p_\mu b_\alpha}\bigg) d_\epsilon b_\alpha + \text{c.c.}=0\label{eq:conservlaw}
\end{align}



In the absence of damping the second term in brackets of Eqn. \ref{eq:conservlaw} vanishes because of the Euler-Lagrange equations. However, in the presence of damping the term in brackets does not vanish due to the augmented E-L equation leading to the general form for the conservation law,
 \begin{align}
    \partial_\mu(\frac{\p \Lag}{\p\partial_\mu b_\alpha}d_\epsilon b_\alpha + \text{c.c})&= 
    \frac{\partial F}{\partial \partial_t b_\alpha}d_\epsilon b_\alpha + \text{c.c.}\label{eq:gs}
\end{align}

It is useful to rewrite this equation using 3+1 notation where j corresponds to z, x, and y, 
\begin{align}
    \partial_t(\frac{\p \Lag}{\p\partial_t b_\alpha}d_\epsilon  b_\alpha) + \partial_j(\frac{\p \Lag}{\p\partial_j b_\alpha}d_\epsilon 
    b_\alpha) + \text{c.c}&= 
    \frac{\partial F}{\partial \partial_t b_\alpha}d_\epsilon b_\alpha + \text{c.c.}\label{eq:gs}
\end{align}

\subsection*{Conservation of Action}
We now look for transformations that leave $\mathcal{L}$ invariant and then find the associated conservation law. We start with 3 phase symmetries that we show are associated with conservation of wave action,
\begin{align*}
    b_0 \mapsto e^{i\theta_2}b_0,  \ &b_1 \mapsto b_1,  \ b_2 \mapsto e^{i\theta_2}b_2,\\
    \end{align*}
    \begin{align*}
    b_0 \mapsto b_0e^{i\theta_1},  \ &b_1 \mapsto e^{i\theta_1} b_1,  \ b_2 \mapsto b_2,\\
\end{align*}
and 
 \begin{align*}
    b_0 \mapsto b_0,  \ &b_1 \mapsto e^{i\theta} b_1,  \ b_2 \mapsto b_2e^{-i\theta}\\
\end{align*}

which for infinitesimal transformations can be written as,

\begin{align*}
    b_0 \mapsto b_0+i\epsilon b_0,
    \ &b_1 \mapsto b_1,  \ b_2 \mapsto b_2+i\epsilon b_2,\\
    \end{align*}

    \begin{align*}
    b_0 \mapsto b_0+i\epsilon b_0,  \ &b_1 \mapsto b_1+i\epsilon b_1,  \ b_2 \mapsto b_2,\\
\end{align*}

and 
 \begin{align*}
    b_0 \mapsto b_0,  \ &b_1 \mapsto b_1+i\epsilon b_0,  \ b_2 \mapsto b_2-i\epsilon b_0\\
\end{align*}

where $d_\epsilon b_\alpha\equiv i\epsilon b_\alpha$. It is straightforward to show that $\mathcal{L}$ is invariant to such transformations.  Applying Eqn. \ref{eq:gs} to the above three symmetries yields the following conservation laws which we write out explicitly in terms of $\partial_t$, $\partial_z$, and $\nabla_\perp$: 

\begin{align}
&\partial_t({\omega_0}|b_0|^2 + \omega_2 |b_2|^2)+\partial_z(k_0|b_0|^2 + \frac{v_e^2k_2}{\Gamma_2} |b_2|^2) +\frac{1}{2i}\nabla_\perp \cdot (b_0^*\nabla_\perp b_0 - b_0\nabla_\perp b_0^* + \frac{v_e^2}{\Gamma_2}(b_2^* \nabla_\perp b_2 - b_2 \nabla_\perp b_2^*)) \nonumber \\
    & = -2\nu_0\omega_0|b_0|^2-2\nu_2\omega_2|b_2|^2 
   \label{eq:mr1} \end{align}
    and
    \begin{align}
&\partial_t({\omega_0}|b_0|^2 + \omega_1 |b_1|^2)+\partial_z(k_0|b_0|^2 + k_1 |b_1|^2) +\frac{1}{2i}\nabla_\perp \cdot (b_0^*\nabla_\perp b_0 - b_0\nabla_\perp b_0^* + b_1^* \nabla_\perp b_1 - b_1 \nabla_\perp b_1^*) \nonumber \\
    & = -2\nu_0\omega_0|b_0|^2-2\nu_1\omega_1|b_1|^2 \label{eq:mr2}
\end{align}
The third Manley-Rowe equation commonly stated in the literature \cite{michel2023introduction} relating $b_1$ and $b_2$ can be obtained by subtracting the second equation from the first, 
\begin{align}
&\partial_t({\omega_1}|b_1|^2 - \omega_2 |b_2|^2)+\partial_z(k_1|b_1|^2 - \frac{v_e^2k_2}{\Gamma_2} |b_2|^2) +\frac{1}{2i}\nabla_\perp \cdot (b_1^*\nabla_\perp b_1 - b_1\nabla_\perp b_1^* - \frac{v_e^2}{\Gamma_2}(b_2^* \nabla_\perp b_2 - b_2 \nabla_\perp b_2^*)) \nonumber \\
    & = -2\nu_1\omega_1|b_1|^2+2\nu_2\omega_2|b_2|^2  \label{eq:mr3}
    \end{align}
    
The RHS of each equation represents sinks to the conserved quantity 
at a rate given by $2\nu_\alpha $. The quantity that is conserved is the one operated on by $\partial_t$. These are $\omega_0 |b_0|^2 + \omega_1 |b_1|^2$, $\omega_0 |b_0|^2 + \omega_1 |b_2|^2$, and $\omega_1 |b_1|^2 - \omega_2 |b_2|^2$.  Terms of the form, $\omega |b|^2$ are well known to be the wave actions (a distinct term from the action $W_{12}$) which are proportional to the density of photons or wave quanta. Thus, these conservation laws are associated with conservation of wave action or the number of photons.  Equations \ref{eq:mr1}-\ref{eq:mr3} are the 3D version of conservation of action and reduce to the well known 1D version of $\nabla_\perp$ is set to zero.

\subsection*{Conservation of Quasi-Energy and Quasi-$z$-momentum}
Additional conservation laws can be obtained from other phase symmetries. Making use of the frequency and wavenumber matching conditions $\omega_0 = \omega_1 + \omega_2, k_{0,0} = k_{1,0} + k_{2,0}$ yields the global symmetries
\begin{align*}
    b_\alpha &\mapsto e^{i\omega_\alpha T}b_\alpha \ \Rightarrow \delta b_\alpha = i\omega_\alpha \epsilon b_\alpha\\
    b_\alpha &\mapsto e^{i k_{\alpha,0} Z}b_\alpha \ \Rightarrow \delta b_\alpha = i k_{\alpha,0
    } \epsilon b_\alpha
\end{align*}
where $d_\epsilon b_\alpha\equiv i
\omega_\alpha\epsilon  b_\alpha$  or  $d_\epsilon b_\alpha\equiv i
 k_{\alpha,0}\epsilon  b_\alpha$  Here ${k}_{\alpha,0} = k_\alpha(0)$ is the wavenumber evaluated at a resonance point where $\phi(0)=0$. Note that  $T,Z$ merely parameterize the two symmetries and do not correspond to the actual time and $z$-coordinate. They can be set to $\epsilon$ for infinitesimal transformations. These symmetries yield conservation laws corresponding to an effective energy and an effective $z$-momentum conservation, which are $\omega_\alpha$ or $k_{\alpha,0}$ times the wave action,  $\omega_\alpha^2|b|^2$ or $k_{\alpha,0} \omega|b|^2$; however as we show in the next section,  these are distinct from the conservation laws for energy and momentum derived from the space symmetries of time and space-translation 
of the Lagrangian density. We thus call these the conservation of quasi-energy and quasi-momentum of the wave. Using $\delta b_{\alpha}=i\omega_{\alpha}\epsilon b_\alpha$ leads to  the conservation of quasi-energy equation 
given by, 
\begin{align}
    &\partial_t
    (\frac{\omega_\alpha^2}{c_\alpha} |b_\alpha|^2)+ \partial_z
    (\frac{\omega_\alpha}{c_\alpha}{k_\alpha} |b_\alpha|^2)+\nabla_\perp \cdot
    \frac{\omega_\alpha v_\alpha^2 }{2c_\alpha i} (b_\alpha^*\nabla_\perp b_\alpha - b_\alpha\nabla_\perp b_\alpha^*))
    = -2
    \frac{\nu_\alpha}{c_\alpha} \omega_\alpha^2 |b_\alpha|^2, 
\end{align}
where, to be rigorously correct, the quasi-energy is $\frac{\omega_\alpha^2}{c_\alpha}|b_\alpha|^2$.

On the other hand, using $\delta b_{\alpha}=ik_{\alpha,0}\epsilon b_\alpha$, leads to  the conservation of quasi-momentum equation  given by,
\begin{align}
    &\partial_t
    (\frac{\omega_\alpha k_{\alpha,0}}{c_\alpha} |b_\alpha|^2)+ \partial_z
    (\frac{k_{\alpha,0}}{c_\alpha}{k_\alpha} |b_\alpha|^2)+\nabla_\perp \cdot
    \frac{k_{\alpha,0} v_\alpha^2 }{ i2c_\alpha} (b_\alpha^*\nabla_\perp b_\alpha - b_\alpha\nabla_\perp b_\alpha^*)
    = -2
    \frac{\nu_\alpha}{c_\alpha} k_{\alpha,0}\omega_\alpha |b_\alpha|^2. 
\end{align}
In both cases, the E-L equations were augmented with the damping function. These conservation laws including loss from damping also reduce to their 1D counterparts when $\nabla_\perp$ is set to zero. 
The symmetries for action, quasi-energy, and quasi-momentum all arise from symmetries of the phases of the waves which are not linearly independent in which case the conservation laws are not linearly independent\cite{michel2023introduction}. 
\subsubsection{Space Symmetries}
In addition to internal symmetries that act only on the field variables themselves,  there are also symmetries regarding transformations on space and time.
Since the action involves integration over time and space then these transformations 
involve variations to the action arising 
from the end points.  
If the Lagrangian depends explicitly on the space variables then the action cannot be invariant unless transformations are made to the part of the Lagrangian that depends explicitly on the space variable.   If the symmetry is parameterized by $\epsilon$ to one of the components of the four vector, or to a linear combination of these components , then the relevant conservation laws can be systematically obtained by differentiating the Lagrangian with respect to each space variable,


\begin{align}
    d_\epsilon \mathcal{L}=
    \bigg(\frac{\mathcal{\p L}}{\p b_\alpha}\p_\epsilon b_\alpha+\frac{\mathcal{\p L}}{\p\p_\mu b_\alpha}\p_\epsilon \p_\mu b_\alpha+c.c.\bigg)+\p_\epsilon \mathcal{L}
\end{align}

Integrating by parts and using the augmented E-L equations that include damping
leads to the following generalized form for the conservation law, 

\begin{align}
    \p_\mu\bigg(\frac{\mathcal{\p L}}{\p\p_\mu b_\alpha}\p_\epsilon b_\alpha+c.c.\bigg) - d_\epsilon \mathcal{L} &=-\partial_\epsilon\mathcal{L}+
    ( \frac{\partial F}{\partial \p_t b_\alpha}\p_\epsilon b_\alpha + c.c.\bigg)\label{eq:modconservlaw}
\end{align}
where for $\partial_\epsilon$ operating on $\mathcal{L}$ the field variables $b_\alpha$ and other spatial variables are kept fixed while for $d_\epsilon$ and $\p_\mu$ only the other spatial variables are kept fixed. 

\section*{Energy Conservation}
As is well known from mechanics,  energy and momentum conservation arises from the invariance of the action to 
space and time translations. In this case 
$x^\nu \mapsto x^\nu+\epsilon$ so that $d_\epsilon = 
\equiv\partial_\nu$ where one is keeping the other spatial (spacetime)  variables fixed but allowing the field variables to vary. We note that we are in flat space time so there is no distinction between covariant and contravariant variables. If the Lagrangian has both damping and an explicit spacetime dependence as in our case, then the energy-momentum tensor conservation equation is modified to,
\begin{align}
    \p_\mu T^{\mu\nu} &= -\p_{\epsilon=\nu} \mathcal{L} + \frac{\partial F}{\partial \p_t b_\alpha}\p_\nu b_\alpha + c.c.
    \label{eq:ls}
\end{align}
where $T^{\mu\nu} = (\frac{\p\mathcal{L}}{\p\p_\mu b_\alpha}\p_\nu b_\alpha + c.c.) - \delta^{\mu\nu} \mathcal{L}$ is the energy-momentum tensor.
Applying Eqn. \ref{eq:ls} with $\nu\rightarrow0$ (i.e., $\nu\rightarrow t$) yields an energy conservation equation with the energy densities and fluxes given by,
\begin{align}
    \mathcal{E}=T^{00} &= (\frac{\p\Lag}{\p \partial_tb_\alpha}\partial_t b_\alpha + \frac{\p\Lag}{\p \partial_tb_\alpha^*}\partial_t b_\alpha^*) - \Lag\label{eq:Epsilon}\\
    \mathcal{S}^i=T^{i0} &= (\frac{\p\Lag}{\p \p_ib_\alpha}\partial_t b_\alpha + \frac{\p\Lag}{\p \p_ib_\alpha^*}\partial_t b_\alpha^*)
    \label{eq:S}
\end{align}
where for the SRS Lagrangian there is no explicit time dependence so $\partial_{\epsilon=t} \mathcal{L}$ vanishes. When evaluated using the E-L equations, the Lagrangian takes the form,
\begin{align}
    \Lag|_{E-L}&= 
    \frac{v_\alpha^2}{4 c_\alpha}\bigg(b_\alpha^* \nabla_\perp^2 b_\alpha + b_\alpha \nabla_\perp^2 b_\alpha^* + 2\nabla_\perp b_\alpha^* \cdot \nabla_\perp b_\alpha \bigg)
    &= 
    \frac{v_\alpha^2}{4 c_\alpha} \nabla_\perp \cdot(b_\alpha^*\nabla_\perp b_\alpha + b_\alpha\nabla_\perp b_\alpha^*)
\end{align}
Inserting this into the expression for the energy density and energy flux (eqns. \ref{eq:Epsilon} and \ref{eq:S}) gives,
\begin{align}
    \mathcal{E} = 
    \frac{\omega_\alpha}{2ic_\alpha}(b_\alpha^* \partial_t b_\alpha - b_\alpha \partial_t b_\alpha^*) &- 
    \frac{v_\alpha^2}{4 c_\alpha}\nabla_\perp \cdot(b_\alpha^*\nabla_\perp b_\alpha + b_\alpha\nabla_\perp b_\alpha^*) \\
    \mathcal{S}^z = 
    \frac{\omega_\alpha v_{g,\alpha}}{2ic_\alpha}(b_\alpha^* \partial_t b_\alpha &- b_\alpha \partial_t b_\alpha^*)\\
    \mathcal{S}^\perp =
    \frac{ v_{\alpha}^2}{2c_\alpha}(\nabla_\perp b_\alpha^* \partial_t b_\alpha &+  \partial_t b_\alpha^* \nabla_\perp b_\alpha)
\end{align}
Since the second term for the energy density is a total divergence, we can absorb the time derivative of this into the energy flux, yielding a simplified form for the conserved quantities,
\begin{align}
    \mathcal{E} =
    \frac{\omega_\alpha}{2ic_\alpha}(b_\alpha^* \partial_t b_\alpha &- b_\alpha \partial_t b_\alpha^*) \label{eq:ed} \\
    \mathcal{S}^z = 
    \frac{\omega_\alpha v_{g,\alpha}}{2ic_\alpha}(b_\alpha^* \partial_t b_\alpha &- b_\alpha \partial_t b_\alpha^*)\\
    \mathcal{S}^\perp = 
    \frac{ v_{\alpha}^2}{4c_\alpha}(\nabla_\perp b_\alpha^* \partial_t b_\alpha &+  \partial_t b_\alpha^* \nabla_\perp b_\alpha - b_\alpha^*\p_t \nabla_\perp b_\alpha - b_\alpha \p_t\nabla_\perp b_\alpha^*)
\end{align}
The full energy conservation equation can therefore be written as,
\begin{align}
    \partial_t \mathcal{E} &+ \partial_z \mathcal{S}^z +\nabla_\perp \cdot \mathcal{S}^\perp = -
    \frac{\omega_\alpha\nu_\alpha}{ic_\alpha}(b_\alpha^*\p_t b_\alpha -\p_t b_\alpha^* b_\alpha)
\end{align}
 and importantly, this is distinct from the conservation of quasi-energy equation (Eqn. 12). 
\section*{$z$-momentum }
    The symmetry according to $z$-momentum invariance is parametrized by $\epsilon=z$ so that $d_\epsilon 
    =\p_z$ and $\p_\epsilon=\p_{\epsilon=z}$. In the inhomogeneous plasma case, this is not an exact symmetry of the Lagrangian; nevertheless, we can still write down an equation for how the $z$-momentum is modified based on Eqns. \ref{eq:modconservlaw} and \ref{eq:ls},
    
\begin{align*}
    \p_\mu\bigg(\frac{\mathcal{\p L}}{\p\p_\mu b_\alpha}\p_z b_\alpha+c.c.\bigg) - \p^z \mathcal{L} =
    \bigg
    ( \frac{\partial F}{\partial \p_t b_\alpha}\p^z b_\alpha + c.c.\bigg) 
\end{align*}
    Hence,  in this case we just need to add the term $-\p_{\epsilon=z} \mathcal{L} = i(b_0 b_1 b_2^* e^{-i\phi(z)} -b_0^* b_1^* b_2 e^{i\phi(z)})\phi'(z)$ to the RHS of the conservation equation.
    The corresponding $z$-momentum density and current are given as follows,
    \begin{align}
    p_z &= (\frac{\p\Lag}{\p \partial_tb_\alpha}\partial_z b_\alpha + \frac{\p\Lag}{\p \partial_tb_\alpha^*}\partial_z b_\alpha^*) \\
    \mathcal{P}^z &= (\frac{\p\Lag}{\p \partial_zb_\alpha}\partial_z b_\alpha + \frac{\p\Lag}{\p \partial_zb_\alpha^*}\partial_z b_\alpha^*) - \Lag\\
    \mathcal{P}^\perp &= (\frac{\p\Lag}{\p \nabla_\perp b_\alpha}\nabla_\perp b_\alpha + \frac{\p\Lag}{\p \nabla_\perp b_\alpha^*}\nabla_\perp b_\alpha^*) 
\end{align}
which yields (when evaluated for the E-L eqns),
\begin{align}
    p_z =
    \frac{\omega_\alpha}{2ic_\alpha}(b_\alpha^* \partial_z b_\alpha &-b_\alpha \partial_z b_\alpha)\\
    \mathcal{P}^z = 
    \frac{\omega_\alpha}{c_\alpha}\bigg(\frac{v_{g,\alpha}}{2i}(b_\alpha^* \partial_z b_\alpha - b_\alpha \partial_z b_\alpha^*) &- \frac{v_\alpha^2}{4\omega_\alpha} \nabla_\perp \cdot(b_\alpha^*\nabla_\perp b_\alpha + b_\alpha\nabla_\perp b_\alpha^*)\bigg) \\
    \mathcal{P}^\perp = 
    \frac{v_\alpha^2}{2c_\alpha}(\nabla_\perp b_\alpha^* \partial_z b_\alpha &+ \partial_zb_\alpha^* \nabla_\perp b_\alpha)
\end{align}
and importantly $\mathcal{P}^z$ has no explicit $z$ dependence. As was the case for the energy density, the transverse divergence in the momentum flux in the $z$ direction
can again be absorbed into the $\perp$-flux,
\begin{align}
    \mathcal{P}^z &= 
    \frac{\omega_\alpha}{c_\alpha}\frac{v_{g,\alpha}}{2i}(b_\alpha^* \partial_z b_\alpha - b_\alpha \partial_z b_\alpha^*)  \\
    \mathcal{P}^\perp &= 
    \frac{v_\alpha^2}{4c_\alpha}(\nabla_\perp b_\alpha^* \partial_z b_\alpha + \partial_zb_\alpha^* \nabla_\perp b_\alpha-b_\alpha^*\nabla_\perp\p_z b_\alpha - b_\alpha\nabla_\perp\p_z b_\alpha^*)
\end{align}
It is now easily demonstrated from the E-L equations that the modification to the conservation of $z$-momentum in the presence of a density gradient and damping is given as follows,
\begin{align*}
    \partial_t p_z + \partial_z \mathcal{P}^z +\nabla_\perp\cdot \mathcal{P}^\perp &= i(b_0 b_1 b_2^* e^{-i\phi(z)} -b_0^* b_1^* b_2 e^{i\phi(z)})\phi'(z)+ 
    \frac{i\omega_\alpha\nu_\alpha}{c_\alpha}(b_\alpha^*\p_z b_\alpha -\p_z b_\alpha^* b_\alpha)
\end{align*}
It should be emphasized that while the damping term is always a momentum sink, the gradient term can either be a source or a sink depending on the sign of $\phi'$. For instance, if $\phi'(z) > 0$, then the wavenumbers of the pump and plasma wave decrease at a faster rate than the rate at which the daughter light wave wavenumber increases due to scattering, leading to a net loss in forward momentum.
\subsubsection*{Total Energy and total Momentum}
We can explain how the energy derived from the Lagrangian based in the invariance of the action due to time translation relates to the quasi-energy derived from conservation of action (phase symmetry). For simplicity, we address this question for the 1d case where there are no transverse gradients. In Fourier space, Eqn. \ref{eq:ed} takes the following form,
\begin{align}
    \mathcal{E}_\alpha &= \frac{\omega_\alpha}{c_\alpha}\int w b_\alpha^*(w')b_\alpha(w)\cos{(w-w')t} dw dw'
\end{align}
By averaging the above equation over timescales $\tau$ much longer than the temporal variation of the envelope, that is $\tau \gg 1 /\delta\omega_\alpha \gg 1 / \omega_\alpha$, one obtains a useful form for the energy density,
\begin{align}
    \langle\mathcal{E}_\alpha\rangle &= \frac{\omega_\alpha}{c_\alpha}\int w b_\alpha^*(w)b_\alpha(w) dw
\end{align}
Thus in Fourier space, one has the following decomposition for the sum of the new expression for the energy and the quasi-energy,
\begin{align*}
    u_{tot,\alpha} = u_{env}+u_{lag} = \frac{\omega_\alpha}{c_\alpha}(\omega_\alpha + \delta\omega)|b_\alpha(\omega_\alpha + \delta\omega,k_\alpha)|^2 
\end{align*}
Heuristically, one can also interpret Eqn. \ref{eq:conservlaw} in real space by writing $b_\alpha = |b_\alpha| e^{i\theta_\alpha}$ \cite{mori1997physics}, where $\theta_\alpha$ is the phase of the wave {\it envelope}. This leads to, 
\begin{align*}
    \mathcal{E}_\alpha \propto \p_t \theta_\alpha |b_\alpha|^2
\end{align*}
and upon recognizing that $\p_t\theta_\alpha = \delta \omega_\alpha$, we have $\mathcal{E}_\alpha \propto \delta \omega_\alpha |b_\alpha|^2$ so that,
\begin{align*}
    \mathcal{E}_{\alpha,total} &= \frac{\omega_\alpha}{c_\alpha} (\omega_\alpha + \p_t \theta_\alpha) |b_\alpha(\omega_\alpha + \p_t\theta_\alpha,k_\alpha)|^2
\end{align*}
The total energy density is the sum of the quasi-energy which is the energy stored in the wave packets themselves and  the additional contribution from the temporal variation of the phase of the wave envelope. A similar interpretation holds for the total $z$-momentum density as well, where the frequencies are replaced with wavenumbers,
\begin{align*}
    p_{z,tot,\alpha} = p_{z,env}+p_{z,action} = \frac{\omega_\alpha}{c_\alpha}(k_\alpha + \delta k)|b_\alpha(\omega_\alpha,k_\alpha + \delta k)|^2 
\end{align*}
while in real space,
\begin{align*}
    p_{z,tot,\alpha} = \frac{\omega_\alpha}{c_\alpha}(k_\alpha +\p_z \theta_\alpha)|b_\alpha(\omega_\alpha,k_\alpha + \delta k)|^2 
\end{align*}
\section*{$\perp$-momentum conservation }
The independence (symmetry) of the Lagrangian to the direction transverse to propagation gives us a momentum conservation law in the two orthogonal directions. The differential is $d_\epsilon = \p_a$ where $a = x,y$.  The $\perp$-momentum density and momentum tensor are given as follows,
    \begin{align}
    p_\perp &= (\frac{\p\Lag}{\p \partial_tb_\alpha}\nabla_\perp b_\alpha + \frac{\p\Lag}{\p \partial_tb_\alpha^*}\nabla_\perp b_\alpha^*) \\
    T^{ab} &=(\frac{\p\Lag}{\p \p_a b_\alpha}\p_b b_\alpha + \frac{\p\Lag}{\p\p_a b_\alpha^*}\p_b b_\alpha^*) - \delta_{ab}\mathcal{L}
\end{align}
The momentum tensor, transverse momentum density, and $\perp$-momentum flux in the $z$-direction can then be simplified as follows,
\begin{align}
    p_\perp = 
    \frac{\omega_\alpha}{2ic_\alpha}(b_\alpha^* \nabla_\perp b_\alpha &-b_\alpha \nabla_\perp b_\alpha^*)\\
    T^{zb} = 
    \frac{\omega_\alpha v_{g\alpha}}{2ic_\alpha}(b_\alpha^* \p_b b_\alpha &- \p_bb_\alpha  b_\alpha^*)\\
    T^{ab} = 
    \frac{ v_\alpha^2}{2c_\alpha}\bigg(\p_ab_\alpha^* \p_b b_\alpha + \p_bb_\alpha \p_a b_\alpha^* &- \frac{1}{2}\delta_{ab}(b_\alpha^* \nabla_\perp^2 b_\alpha + b_\alpha \nabla_\perp^2 b_\alpha^*) - \delta_{ab}\nabla_\perp b_\alpha^* \cdot \nabla_\perp b_\alpha \bigg)\nonumber\\
\end{align}
such that the $\perp$-momentum conservation equation (with damping) becomes,
\begin{align*}
    \p_t p_\perp + \p_z T^{zb}+ \p_a T^{ab} &= -\sum\frac{\omega_\alpha}{ic_\alpha}(b_\alpha^*\nabla_\perp b_\alpha - b_\alpha \nabla_\perp b_\alpha^*)
\end{align*}
\section*{Angular Momentum Conservation}
Finally, invariance under rotations around the axial
axis yields an angular momentum conservation law identical to OAM conservation. In this case $d_\epsilon = \p_\phi$ where $\p_\phi = x \p_y - y \p_x$.
The angular momentum density and flux is given by:
\begin{align}
    \L_z &= (\frac{\p\Lag}{\p \partial_tb_\alpha}\p_\phi b_\alpha + \frac{\p\Lag}{\p \partial_tb_\alpha^*}\p_\phi b_\alpha^*) \\
    \mathcal{M}^z &= (\frac{\p\Lag}{\p \partial_zb_\alpha}\p_\phi b_\alpha + \frac{\p\Lag}{\p \partial_zb_\alpha^*}\p_\phi b_\alpha^*) \\
    \mathcal{M}^a &= (\frac{\p\Lag}{\p \p_a b_\alpha}\p_\phi  b_\alpha + \frac{\p\Lag}{\p \p_a b_\alpha^*}\p_\phi b_\alpha^*) - (y \delta_{a x} - x \delta_{a y})\Lag\\
    &= \epsilon_{zbc}x^b T^{ac} \nonumber
\end{align}
The density and fluxes can be written explicitly as,
\begin{align*}
    L_z &=  
    \frac{\omega_\alpha}{2ic_\alpha}(b_\alpha^* \p_\phi b_\alpha -b_\alpha \p_\phi b_\alpha)\\
    \mathcal{M}^z &= 
    \frac{\omega_\alpha v_{g,\alpha}}{2ic_\alpha}(b_\alpha^* \p_\phi b_\alpha - b_\alpha \p_\phi b_\alpha^*)
\end{align*}
Finally, the damping rate that appears on the RHS of the conservation equation is,
\begin{align*}
    \mathcal{R} &= -
    \frac{\omega_\alpha}{ic_\alpha}(b_\alpha^*\partial_\phi b_\alpha - b_\alpha \partial_\phi b_\alpha^*)
\end{align*}
and the conservation law for OAM can now be written as,
\begin{align*}
    \p_tL_z+\p_z \mathcal{M}^z+\p_a\mathcal{M}^a=\mathcal {R}
\end{align*}
We stress that such a conservation law holds for SRS or SBS growing from noise or from a seed. It also applies to a multi-speckled beam. 
\subsection*{OAM for light waves}
As described in \cite{allen1992orbital} the above expressions for the OAM of a light wave can be derived from the definition of angular momentum in the electromagnetic fields, $\mathbf{L_{EM}}=\mathbf{r}\times \mathbf{P}$ where $\mathbf{P}=
\frac{\mathbf{E} \times  \mathbf{B}}{4\pi c}$.
Recall that in the slowly varying envelope approximation the vector potential in the Coulomb gauge for a linearly polarized light wave is written as:
\begin{align*}
    \mathbf{A}(\mathbf{x}) &= \mathbf{\hat{x}}u(x,y,z)e^{i(kz- \omega t)}
\end{align*}
In terms of the envelopes, the momentum density for a light wave is $\mathbf{P} = \frac{1}{2}(\mathbf{E}^*\times \mathbf{B} +c.c.)$ which in terms of the vector potential can be written as $\frac{i}{2} \omega(\nabla \mathbf{A}^* \cdot \mathbf{A} - c.c.)$. Rewriting this in terms of the envelope leads to \cite{allen1992orbital}

\begin{align*}
    \mathbf{P} &=\frac{\omega}{2i}(u^*\nabla u - u\nabla u^*)+\omega k|u|^2\mathbf{\hat{z}}
\end{align*}
The $z$-angular momentum density is then:
\begin{align*}
    L_z &= (\mathbf{r}\times \mathbf{P})\cdot \mathbf{\hat{z}} = x P_y -yP_x\\
    &= \frac{\omega}{2i}(u^*\p_\phi u - u\p_\phi u^*)
\end{align*}
which agrees precisely with the angular momentum derived from the Lagrangian. We note that it is straightforward to extend this to allow for circularly polarized light and the associated spin angular momentum\cite{allen1992orbital}. 
\subsection*{OAM for plasma waves}
The orbital angular momentum of the plasma waves requires a slightly more involved treatment in order to see its physical origin. Following \cite{sarah}, the fluid equation describing the density perturbation reads, in the electrostatic approximation:
\begin{align*}
     \p_t \mathbf{v}_{e,\perp} &= \frac{e}{m_e}\nabla_\perp \varphi - \frac{\gamma_ev_t^2}{m_e}\nabla_\perp\tilde{n}_e
\end{align*}
where $n_e$ is the electron density and $\gamma_e=3$ is the coefficient in the adiabatic gas law for the equation of state for the plasma electrons. The electrostatic potential $\varphi$ can be solved via Gauss' law:
\begin{align*}
    \nabla^2\varphi = -k^2\varphi + \nabla_\perp^2\varphi = -\frac{4\pi e}{n_{e,0}} \tilde{n}_e
\end{align*}
solving for $\varphi$ and Taylor expanding to leading order in the transverse derivatives yields
\begin{align*}
   \varphi \approx -\frac{4\pi e}{n_{e,0}}\tilde{n}_e
\end{align*}
and inserting this into the fluid equation leads to,
\begin{align*}
    -i\omega \mathbf{v}_{e,\perp} &= -\frac{\omega_p^2}{k^2}\nabla_\perp\tilde{n}_e - \frac{\gamma_ev_t^2}{m_e}\nabla_\perp\tilde{n}_e\\
    &= -\frac{1}{k^2}\omega^2 \nabla_\perp \tilde{n}_e
\end{align*}
where we have identified $\omega^2 = \omega_p^2 + \frac{\gamma_e v_t^2}{m_e} k^2$.
The time-averaged perpendicular momentum density  in normalized units is therefore,
\begin{align*}
    \tilde{\mathbf{P}}_{\perp,e} \equiv\frac{\langle \mathbf{P}_{\perp,e}\rangle}{n_{e,0}m_e c}  &= \frac{1}{4c}(\tilde{n}_e^*  \mathbf{v}_{\perp,e} + \tilde{n}_e  \mathbf{v}_{\perp,e}^*)\\
    &= \frac{\omega}{4ick^2} (\tilde{n}_{e}^*\nabla_\perp \tilde{n}_e-\tilde{n}_{e}\nabla_\perp \tilde{n}_e^*)
\end{align*}
Recall that $\Gamma_2 = c^2 k^2/\omega_p^2$ so we have:
\begin{align*}
    \frac{\langle \mathbf{P}_{\perp,e}\rangle}{n_{e,0}m_e c}  
    &= \frac{1}{4i\Gamma_2}\frac{\omega}{\omega_p} (\tilde{n}_{e}^*\frac{c}{\omega_p}\nabla_\perp \tilde{n}_e-\tilde{n}_{e}\frac{c}{\omega_p}\nabla_\perp \tilde{n}_e^*)
\end{align*}
From this, one can see that the $z$-component of the angular momentum for the electrons agrees with the expression from the Lagrangian when written in normalized units.

\subsubsection{OAM Index of an arbitrary wave}
The conservation of OAM equation can be used to investigate under what circumstances the conservation of $L_z$ can be described as an $\ell$ matching condition, in analogy with how the conservation of energy and momentum are related to
$\omega$ and $k$ matching conditions . In \cite{allen1992orbital} it was shown that the OAM per photon of a pure Laguerre Gaussian mode, where the envelope is proportional to $e^{i\ell\phi}$, is 
$\ell \hbar$.  We can use the conservation law for $\L_z$ to define an effective OAM index for any envelope\cite{sarah}. 
The effective OAM index for wave $j$ is the ratio of the 
angular momentum density integrated over $d\mathbf{x}_{\perp}$ to the total 
action integrated over $d\mathbf{x}_{\perp}$,
\begin{equation}
\label{eq:ell-def}
    \bar{\ell}_{j} = \frac{-\frac{i\omega_j}{2}\int\left( b_j^{*} \partial_{\phi} b_j - b_j\partial_{\phi} b_j^{*}\right)d\mathbf{x_{\perp}}}{\omega_j\int |b_j|^{2} d\mathbf{x_{\perp}}}.
\end{equation}
In general, we can expand an arbitrary envelope, $b$, as a superposition of OAM states each with integer $\ell$,
\begin{equation}
    b(\mathbf{x_{\perp}}) = \sum_{\ell=-\infty}^{\infty} b_{\ell}(r) e^{i\ell\phi}
\end{equation}
where the effective OAM index, $\bar{\ell}$, (equation \ref{eq:ell-def}) is the weighted average,
\begin{equation}
    \bar \ell = \frac{\sum_{\ell} \ell |b_{\ell}|^{2}}{\sum_{\ell'} |b_{\ell'}|^{2}}
\end{equation}
For a pure Laguerre Gaussian envelope in vacuum there is a term $e^{i\ell_{0}\phi}$, the index $\ell_{0}$ is recovered. Note that the action of the wave with frequency $\omega_j$ is defined as $\mathbf{a}_j\equiv\omega_{j}\sum_{\ell'}|b_{\ell'}|^2$ so the OAM of the wave is $\bar\ell_j\mathbf{a}_j$.

In the literature the concept of $\ell$ matching has been invoked with respect to SRS \cite{mendoncca2009stimulated}. This can be easily argued if each wave is a pure OAM Laguerre-Gaussian mode.  Consider $b_0\propto e^{il_0\phi}$, $b_1\propto e^{il_1\phi}$, and $b_2\propto e^{il_2\phi}$ then in order for each wave to be resonantly driven then the RHS of each wave equation (Eqns. \ref{eq: 0} to \ref{eq:2}) must have the same $\ell$ as that on the LHS $\ell_j$, which leads directly to $\ell_0=\ell_1+\ell_2$.

However, this implies that the $\ell_{j}$ number of each wave cannot change. The same assumption is true for $\omega$ and $k$ matching, where it is implicitly assumed that the $\omega_{j}$ and $k_{j}$ of each wave does not change. However, these quantities are not part of the envelope, while OAM is an intrinsic property of the envelope itself. 
Thus, the OAM selection rule or $\ell$ matching is more subtle than for $\omega$ and $k$. 

To clarify the above discussion, it is illustrative to see how $\omega$ matching follows from the conservation of energy equation. The same arguments apply to $k$ matching and the conservation of momentum equation. Consider the conservation of energy equation after integrating over $d\mathbf{ x_{\perp}}$,
\begin{equation}
    \partial_{t}\left( \omega_{0}\bar a_{0} + \omega_{1}\bar a_{1} + \omega_{2} \bar a_{2} \right) +  \partial_{z}\left( v_{g0}\omega_{0}\bar a_{0} + v_{g1}\omega_{1}\bar a_{1} + v_{g2}\omega_{2}\bar a_{2} \right) = 0
\end{equation}
where $\bar a\equiv \int d\mathbf {x_{\perp}} \mathbf{a}$ and $\mathbf{a}_{2}$ is defined as $\frac{1}{\Gamma_{2}}|b_{2}|^{2}$. This equation can be rewritten as
\begin{multline}
    \bar a_{0} (\partial_{t} + v_{g0}\partial_{z}) \omega_{0} + \bar a_{1} (\partial_{t} + v_{g1}\partial_{z}) \omega_{1} + \bar a_{2} (\partial_{t} + v_{g2}\partial_{z}) \omega_{2} \\ + \omega_{0} (\partial_{t}\bar a_{0} + \partial_{z}v_{g0}
\bar a_{0}) + \omega_{1} (\partial_{t}
\bar a_{1} + \partial_{z}v_{g1}\bar a_{1}) + \omega_{2} (\partial_{t}\bar a_{2} + \partial_{z}v_{g2}\bar a_{2}) = 0
\end{multline}
which can then be rewritten as 
\begin{multline}
    \bar a_{0} (\partial_{t} + v_{g0}\partial_{z}) \omega_{0} + \bar a_{1} (\partial_{t} + v_{g1}\partial_{z}) \omega_{1} + \bar a_{2} (\partial_{t} + v_{g2}\partial_{z}) \omega_{2} \\ + (\omega_{0}-\omega_{1}-\omega_{2})(\partial_{t}\bar a_{0}+\partial_{z}v_{g0}\bar a_{0})=0
\end{multline}
if the conservation of action equations are used. This equation is satisfied if $\omega$ matching, $\omega_{0}-\omega_{1}-\omega_{2}=0$, and each frequency is constant. Thus, $\omega$ matching follows from energy conservation only if each $\omega_j$ does not change. Analogous conclusions can be made for $k$ matching by combining the momentum and conservation of action equations.

Following the same arguments,  we can combine conservation of OAM with the conservation of action equations leading to,
\begin{multline}
    \bar a_{0} (\partial_{t} + v_{g0}\partial_{z}) \bar{\ell}_{0} + \bar a_{1} (\partial_{t} + v_{g1}\partial_{z})  \bar{\ell}_{1} + \bar a_{2} (\partial_{t} + v_{g2}\partial_{z})  \bar{\ell}_{2} \\  + (\bar{\ell}_{0}-\bar{\ell}_{1}-\bar{\ell}_{2})(\partial_{t}\bar a_{0}+\partial_{z}v_{g0}\bar a_{0})=0
\end{multline}
which makes clear that $\bar{\ell}$ matching is only true if each $\bar{\ell}$ remains constant. This is generally not true, so $\bar{\ell}$ matching is not an absolute concept in SRS while the conservation equation for $M_z$ is absolutely true.

\section{Extending the Lagrangian}

To properly study some problems of interest the starting equations might need to be modified. For example, higher order corrections to the paraxial ray equations,  including full temporal operators, and the addition of nonlinear frequency shifts to the EPW or IAW might be needed. In this section, we show how the Lagrangian can be straightforwardly modified to include some of these effects. We also mention how the conservation laws would be augmented for the full temporal operator. 

\subsection*{Strongly coupled Raman and Brillouin: Beyond the envelope approximation }
There are situations of interest, e.g.,  Brillouin amplification\cite{Alves_2021}, where SRS or SBS occurs in what is referred to as the 
strongly coupled regime \cite{forslund1975theory}. Furthermore, current XFEL facilities operate in a regime where the instability is equivalent to SRS backscatter in the strongly coupled regime when viewed in the frame of the electron beam (the wiggler is a quasi-electromagnetic wave in this frame)\footnote{It can be shown from the formula for the Pierce parameter that the gain for an XFEL is $G=\frac{\sqrt3}{2}[\frac{1}{4}|b_0|^2k_w'\omega _p'^2/c^2]^{1/3}L$ where $b_0$ is the normalized strength of the wiggler, and $k_0'$, $\omega_p'$, and L are the wavenumber of the wiggler, the plasma frequency of the beam, and the length of the wiggler in the beam's frame; which is identical for SRS where the wiggler is replaced by an incoming laser.}. In this regime the growth rate and frequency of the EPW(or IAW) greatly exceeds the linear frequency. This implies that the envelope does not evolve slowly compared to $\omega_2$. Under these conditions,
\ref{eq:2} is modified to
\begin{align}
    (\frac{i[\p_t^2-u_e^2\p_z^2]}{2} + \omega_e[\p_t+v_{g,e} \p_z] - \frac{iv_e^2}{2}\nabla_\perp^2)b_2 &= ic_2 \tilde{b}_0\tilde{b}_1^*e^{i\phi(z)}
\end{align}
This implies that one must augment the EPW (or IAW) dynamical term of the Lagrangian which is not difficult to accomplish. For simplicity, however, here we limit the discussion to a regime where the $\p_t^2$ dominates the temporal part and the $\p_z^2$ is negligible for the spatial part, then we are left with,

\begin{align}
    (\frac{i\p_t^2}{2} + \omega_ev_{g,e} \p_z - \frac{iv_e^2}{2}\nabla_\perp^2)b_2 &= ic_2 \tilde{b}_0\tilde{b}_1^*e^{i\phi(z)}
\end{align}

This equation for the EPW (or IAW) can be recovered if the dynamic term of the Lagrangian is replaced with, 
\begin{align}
    \mathcal{L}_{e,dyn} &= \frac{1}{2c_e}\p_tb_2^*\p_tb_2
\end{align}
With this modification, the energy, momentum and angular momentum densities can be recomputed. Additionally, the Lagrangian evaluated at the E-L equations picks up a total time derivative:
\begin{align*}
    \mathcal{L}|_{E-L} &= \sum_\alpha \frac{v_\alpha^2}{4 c_\alpha} \nabla_\perp \cdot(b_\alpha^*\nabla_\perp b_\alpha + b_\alpha\nabla_\perp b_\alpha^*) + \frac{1}{2c_e}\p_tb_2^*\p_tb_2 - \frac{1}{4c_e}\p_t(b_2^*\p_tb_2 + \p_tb_2^*b_2)
\end{align*}
The EPW energy density is modified to,
\begin{align}
    \mathcal{E}_e =  \frac{1}{2}\p_tb_2^*\p_tb_2+\frac{1}{4}(b_2^* \partial_t^2 b_2 + b_2 \partial_t^2 b_2^*) &-  \frac{v_e^2}{4 c_e}\nabla_\perp \cdot(b_2^*\nabla_\perp b_2 + b_2\nabla_\perp b_2^*) 
\end{align}
The EPW momentum density and z-flux are therefore both modified as,
\begin{align*}
    p_{z,e} = \frac{1}{2ic_e}(\p_tb_2^* \partial_z b_2 &-\p_zb_2^* \p_t b_2)\\
    \mathcal{P}_{z,e} = -\bigg(\frac{\omega_e}{ic_e}\bigg(\frac{v_{g,e}}{2}(b_2^* \partial_z b_2 - b_2 \partial_z b_2^*) &+ \frac{iv_e^2}{4\omega_e} \nabla_\perp \cdot(b_2^*\nabla_\perp b_2 + b_2\nabla_\perp b_2^*)\bigg)\\
    & + \frac{1}{2c_e}\p_tb_2^*\p_tb_2 + \frac{1}{4c_e}(b_2^*\p_t^2b_2 + b_2\p_t^2b_2^*)\bigg) \\
\end{align*}
the transverse momentum is modified as, 
\begin{align*}
    p_{\perp,e} =  \frac{1}{2ic_e}(\p_tb_2^* \nabla_\perp b_2 &-\p_t b_2 \nabla_\perp b_2^*)\\
    T_{ab,e} =  \frac{ v_e^2}{2c_e}\bigg(\p_ab_2^* \p_b b_2 + \p_bb_2 \p_a b_2^* &- \frac{1}{2}\delta_{ab}(b_2^* \nabla_\perp^2 b_2 + b_2 \nabla_\perp^2 b_2^*) - \delta_{ab}\nabla_\perp b_2^* \cdot \nabla_\perp b_2 \bigg)\\
    &- \delta_{ab}\bigg( \frac{1}{2c_e}\p_tb_2^*\p_tb_2 + \frac{1}{4c_e}(b_2^*\p_t^2b_2 + b_2\p_t^2b_2^*) \bigg)
\end{align*}
and the OAM is modified  as, 
\begin{align*}
    \ell_{z,e} &= \frac{1}{2c_e}(\p_t b_2^*\p_\theta b_2 + \p_t b_2 \p_\theta b_2^*)\\
    \mathcal{M}_{z,e} &= \frac{1}{2c_e}(\p_z b_2^*\p_\theta b_2 + \p_z b_2 \p_\theta b_2^*)\\
    \mathcal{M}_{a,e} &= \epsilon_{zbc}x^b T_e^{ac}
\end{align*}
If more corrections to the paraxial ray approximation are needed then the Lagrangian can be modified appropriately. 
\subsection*{Nonlinear frequency shifts}
In SRS and SBS sometimes nonlinear corrections to the frequency and damping of the electrostatic waves is important. For example, in inflationary SRS \cite{VuPhysRevLett} damping rate can decrease to zero and there are nonlinear changes to the frequency. Furthermore, nonlinear frequency shifts can lead to an autoresonance phenomena in a density gradient\cite{Wurtele_autoresonance_PhysPlasmas}\cite{Chapman_autoresonance_PRL}. 

In nonlinear EPWs with sufficiently large $k\lambda_d$  the nonlinear frequency shift scales as the amplitude of the wave to the $1/2$ power \cite{MoralesPhysRevLett}, and is due to kinetic effects from trapped particles. In reality the operator for this nonlinearity is nonlocal in space and time as trapped electrons move from one point in space to another.  There can also be nonlinear frequency shifts due to harmonic generation and these scale as the amplitude squared\cite{WinjumPhysPlasmas}. There can also be nonlinear frequency shifts for IAWs due to trapped particles,  harmonics, or multi-species\cite{BergerPhysPlasmas}\cite{ChapmanPhysRevLett}. We show it is straightforward to include local nonlinear frequency shift of arbitrary power into the Lagrangian. Even if such frequency shift is not quantitatively accurate, the resulting equations will include the effects at a phenomenological level to permit studies of autoresonance and detuning.  

In the presence of such a nonlinear frequency shift the envelope equation is modified to
\begin{align}
    \omega_2(\partial_t + \nu_2 +v_{g,2} \partial_z  -\frac{iu_2^2}{2\omega_2}\nabla_{\perp}^2 - i\beta |b_2|^\gamma - 
    \frac{\partial_z v_{g,2}}{2} )b_2 &= -ic_2 b_0b_1^*e^{i\phi(z)} \label{eq:ar}\\
\end{align}
where $\gamma = 2, 1/2$. 
The appropriate term to add to the Lagrangian is the following expression,
\begin{align*}
    \delta \mathcal{L}_{ar} = \frac{\omega_2}{c_2}\frac{2\beta}{\gamma + 2}|b_2|^{\gamma + 2}
\end{align*}
The Lagrangian evaluated at the E-L equations is now modified to, 
\begin{align*}
    \Lag|_{E-L,2}
    &= 
    \frac{v_2^2}{4 c_2} \nabla_\perp \cdot(b_2^*\nabla_\perp b_2 + b_2\nabla_\perp b_2^*) + \frac{\omega_2 \beta}{c_2}\frac{4+\gamma}{2+\gamma}|b_2|^{\gamma + 2}
\end{align*}
The energy density and momentum fluxes for the 2nd wave when it is an EPW are therefore modified as follows,
\begin{align}
    \mathcal{E}_e &= 
    \frac{\omega_2}{2ic_2}(b_2^* \partial_t b_2 - b_2 \partial_t b_2^*)-\frac{v_2^2}{4 c_2} \nabla_\perp \cdot(b_2^*\nabla_\perp b_2 + b_2\nabla_\perp b_2^*) -\frac{\omega_2 \beta}{c_2}\frac{4+\gamma}{2+\gamma}|b_2|^{\gamma + 2}\\
    \mathcal{P}_{z,2} &= -\frac{\omega_2}{c_2}\bigg(\frac{v_{g,2}}{2i}(b_2^* \partial_z b_2 - b_2 \partial_z b_2^*) + \frac{v_e^2}{4\omega_e} \nabla_\perp \cdot(b_2^*\nabla_\perp b_2 + b_2\nabla_\perp b_2^* ) + \beta\frac{4+\gamma}{2+\gamma}|b_2|^{\gamma + 2}\bigg)\\
     T_{ab,2} &=  \frac{ v_2^2}{2c_2}\bigg(\p_ab_2^* \p_b b_2 + \p_bb_2 \p_a b_2^* - \frac{1}{2}\delta_{ab}(b_2^* \nabla_\perp^2 b_2 + b_2 \nabla_\perp^2 b_2^*) - \delta_{ab}\nabla_\perp b_2^* \cdot \nabla_\perp b_2 \bigg)\\
    & - \delta_{ab}\frac{\omega_2 \beta}{c_2}\frac{4+\gamma}{2+\gamma}|b_2|^{\gamma + 2}
\end{align}
Last, the modification to the $x$ and $y$-components of the z-angular momentum flux are given as follows,
\begin{align*}
    \delta\mathcal{M}_{y,2} &= - x\frac{\omega_2 \beta}{c_2}\frac{4+\gamma}{2+\gamma}|b_2|^{\gamma + 2}\\
    \delta\mathcal{M}_{x,2} &= y\frac{\omega_2 \beta}{c_2}\frac{4+\gamma}{2+\gamma}|b_2|^{\gamma + 2}
\end{align*}
\section*{Summary}
A Lagrangian formulation of 3-wave coupled equations that describe SRS and SBS backscatter within the paraxial ray approximation that also accounts for a 
slowly-varying density gradient is presented. Recasting the three-wave equations for the envelopes of each wave in terms of a Lagrangian provides a systematic procedure based on Noether's theorem for identifying conservation laws for the system. We show how to augment the conservation laws to include sinks from damping. Some of these 3D conservation laws reduce to well known 1D versions of the Manley-Rowe  relations (conservation of action) and conservation of energy and momentum. However, the 3D versions are new and additional conservation laws (even in 1D) are found for the energy, momentum, and OAM of the wave envelopes. These conservation laws hold for SRS and/or SBS arising from a diffraction limited  or a multi-speckled pump with or without bandwidth, from noise, or from a well defined seed. Using the OAM equation,  we define an OAM index for each wave and discuss under what conditions there is matching between the OAM indices. While interesting in their own right, these conservation laws also provide equations to test simulation codes, such as pF3D, that are based on the three coupled wave equations. Modifications to the Lagrangian to include strongly coupled SRS and SBS and nonlinear frequency shifts to the electrostatic waves are also given.  Finally, we note that the Lagrangian itself can also have other uses. For example, trial functions that have well defined transverse properties/parameters that depend on (z,t)  can be substituted into the action and then integration over the transverse variables can be carried out. Variation of this reduced Lagrangian provides 1D Euler-Lagrange equations for the parameters. In addition,  trial functions that describe convective or absolute growth can be  substituted into the Lagrangian to aid in determining thresholds. We leave these later ideas for future work.

We acknowledge useful discussions with Prof. Paulo Alves, Dr. Mik
hail Belyaev,  and Maria Almanza.  This work was supported under DOE grant number DE-NA-0004131,  LLE subcontracts SUB000001031 and SUB00000211, and the LLNL ACT Up program. The work of J.P.P. was supported by the Department of Energy National Nuclear Security Administration under Award Number DE-NA0004144. The work of D.E.H.  was performed under the auspices of the U.S. Department of Energy by Lawrence Livermore National Laboratory under Contract DE-AC52-07NA27344.

\bibliography{references} 

\end{document}